\documentclass[%
 reprint,
nofootinbib,
 amsmath,amssymb,
 aps,
]{revtex4-1}
\usepackage{comment}
\usepackage{graphicx}
\usepackage{dcolumn}
\usepackage{bm,color}
\usepackage{hyperref}

\newtheorem{remark}{Remark}

\begin{document}

\preprint{APS/123-QED}

\title{Geometric Brownian motion with intermittent entries and exits}

\author{Suvam Pal}

\email{suvamjoy256@gmail.com}
\affiliation{Physics and Applied Mathematics Unit, Indian Statistical Institute, 203 B.T. Road, Kolkata, India}%

\author{Viktor Stojkoski}%
\email{vstojkoski@eccf.ukim.edu.mk} 
\affiliation{Faculty of Economics, Ss.~Cyril and Methodius University in Skopje, 1000 Skopje, Macedonia}
\affiliation{Center for Collective Learning, University of Corvinus, Budapest, Hungary}%

\author{Arnab Pal}%
\email{arnabpal@imsc.res.in}
\affiliation{The Institute of Mathematical Sciences, CIT Campus, Taramani,
Chennai 600113, India}
\affiliation{Homi Bhabha National Institute, Training School Complex, Anushakti Nagar, Mumbai, Maharashtra 400094, India}%

\author{Trifce Sandev}%
\email{trifce.sandev@manu.edu.mk}
\affiliation{Research Center for Computer Science and Information Technologies, Macedonian Academy of Sciences and Arts, Bul. Krste Misirkov 2, 1000 Skopje, Macedonia}
\affiliation{Institute of Physics, Faculty of Natural Sciences and Mathematics, Ss.~Cyril and Methodius University, Arhimedova 3, 1000 Skopje, Macedonia}
\affiliation{Department of Physics, Korea University, Seoul 02841, Korea}%


\date{\today}

\begin{abstract}
We study a generalized geometric Brownian motion framework that incorporates both entries of new units and exit mechanisms for the current population, extending earlier stochastic resetting models where these rates are treated as identical. The model captures realistic features observed in many economic observables, which can be explained as market-driven firm entries/exits, worker inflow/outflow, and income growth/loss. This model is not conservative and, despite the asymmetry in the entry and exit rates, we find that the system eventually relaxes to a stationary distribution. Moreover, our analysis reveals three distinct dynamical regimes in the moments of the distribution, arising from the interplay between volatility, drift, entry, and exit rates. We further derive the survival probability and the mean first-passage time associated with the observed variable reaching certain threshold under the competing entry-exit processes. Interestingly, we identify an optimal exit rate that minimizes the mean first-passage time, providing insights into how entry and exit policies can influence the outcome of the system. These results should be useful for understanding the long-run behavior of economic systems in which growth, volatility, entry, and exit jointly shape the evolution of heterogeneous units.
\end{abstract}

\maketitle


\section{Introduction}
Geometric Brownian motion (GBM) is one of the central stochastic processes used to model multiplicative growth in natural and social systems~\cite{redner1990random}. By construction, it preserves positivity and produces log-normal statistics, which made it the standard model in finance for asset prices~\cite{black1973pricing,Merton1974} and a natural benchmark in economics for income~\cite{aitchison1954criteria,gabaix2016dynamics,stojkoski2022income,jolakoski2023first}, wealth~\cite{berman2020wealth,stojkoski2022ergodicity}, and population dynamics~\cite{DeLauro2014,dennis1991estimation,engen2007stochastic,lande2003stochastic,kemp2022statistical}. The same structure also appears in diverse settings such as diffusion in heterogeneous media and turbulent transport~\cite{fuchs2022instantons,friedrich1997description,baskin2004superdiffusion,sandev2020hitting}, and many other systems driven by multiplicative noise~\cite{limpert2001log}.

Despite its broad applicability, most existing formulations of GBM describe closed systems, where the number of trajectories (or agents) is fixed~\cite{peters2013ergodicity,peters2019ergodicity}. In many empirical settings, however, the relevant systems are inherently open: units continuously enter and exit the population. This is well documented in firm dynamics, where entry, growth, and exit are central empirical regularities~\cite{DunneRobertsSamuelson1988,DunneRobertsSamuelson1989,Hopenhayn1992}; in credit markets, where loan origination and default occur continuously over time~\cite{QuerciaStegman1992,MayerPenceSherlund2009,FooteGerardiWillen2018}; and in entrepreneurial ecosystems, where startup formation and failure are pervasive features of the underlying dynamics~\cite{OECD2023EntrepreneurialEcosystems,WorldBankEntrepreneurialEcosystems}. In such systems, the interplay between multiplicative growth and turnover is central in determining the stationary distribution, the evolution of moments, and first-passage properties, as shown in this paper.

\begin{figure}[h]
    \centering
    \includegraphics[width=\linewidth]{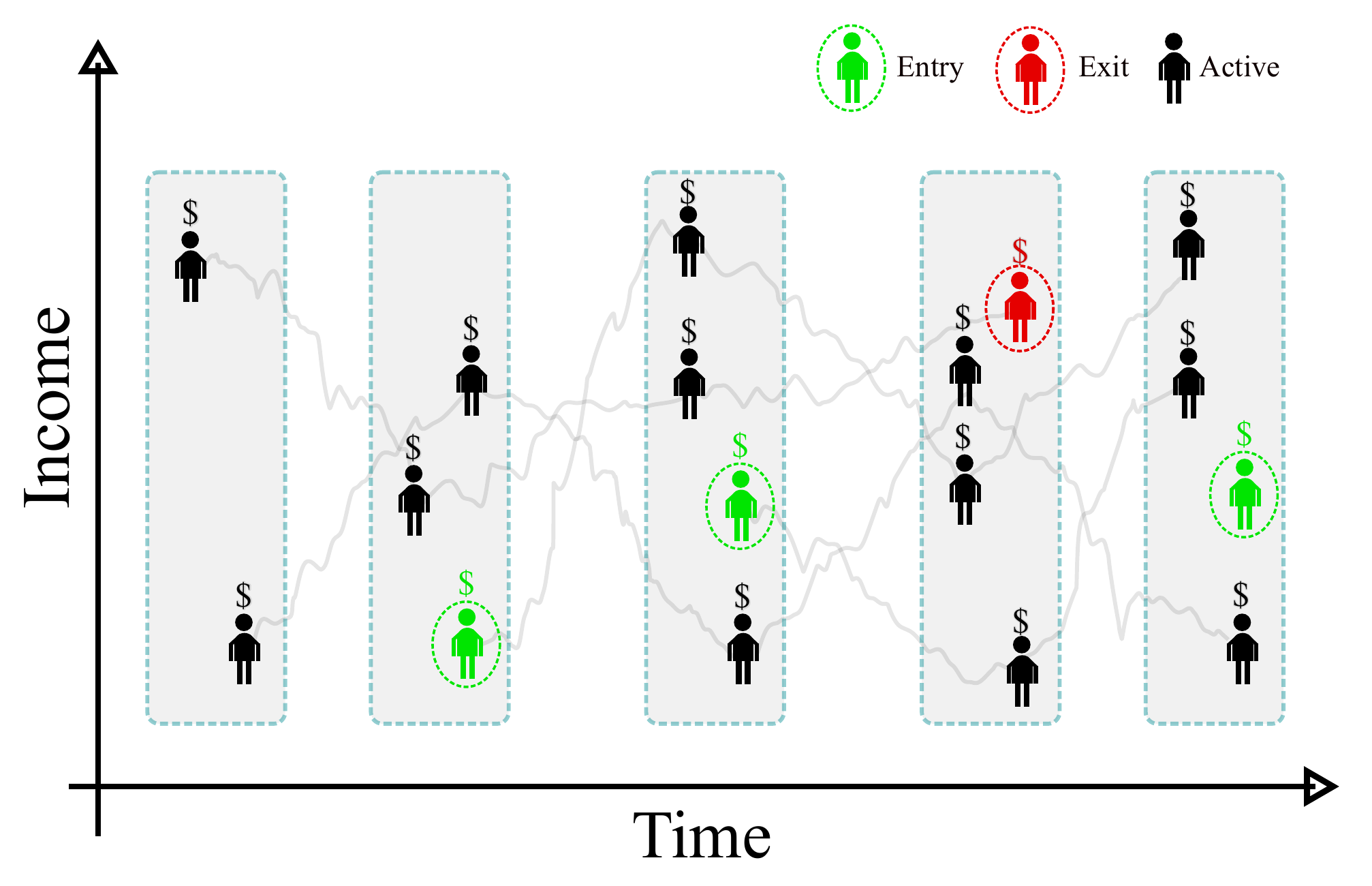}
    \caption{Schematic illustration of income dynamics governed by geometric Brownian motion in the presence of entry and exit mechanisms. At the initial time of observation, the system consists of two active individuals with specified income levels that evolve stochastically over time (background trajectories). Entry and exit events dynamically alter the population size, thereby modifying the aggregate income distribution over time. Using this set-up, we study both the steady state properties as well as threshold-crossing time statistics.}
    \label{schematic}
\end{figure}

This distinction matters because many of the heavy-tailed size distributions observed in the data arise precisely in settings with ongoing turnover. Firm sizes are a canonical example: the empirical distribution is highly skewed, and in large samples its upper tail is often close to Zipf's law~\cite{Axtell2001}. Related work has shown that proportional growth, selection, and entry are central ingredients in accounting for such distributions~\cite{Luttmer2007}. More generally, the literature on double-Pareto and double-Pareto-lognormal distributions has emphasized that birth-death structure and heterogeneous ages can generate power-law behavior in one or both tails~\cite{Reed2001,Reed2003,ReedJorgensen2004}. Yet these insights are usually developed either in descriptive distributional terms or in models tailored to a specific application, rather than in a unified continuous-time GBM framework with separate entry and exit mechanisms.

A natural way to extend GBM to open systems is to introduce entries and exits, whereby new trajectories enter at a characteristic scale and existing ones exit at a constant rate (see Fig. (\ref{schematic})). The closest related literature to this idea is that on stochastic resetting, in which trajectories are intermittently returned to a reference state \citep{EvansMajumdar2011,evans2014diffusion,pal2015diffusion,EvansMajumdarSchehr2020,palconditional2025}. Resetting has proved especially fruitful because it generates non-equilibrium stationary states~\cite{EvansMajumdar2011,pal2015diffusion,EvansMajumdarSchehr2020,pal2016diffusion} and fundamentally reshapes first-passage properties, often yielding finite and even optimal search times \citep{EvansMajumdar2011,PalReuveni2017,EvansMajumdarSchehr2020,reuveni2016optimal,ray2021mitigating}. See reviews \cite{EvansMajumdarSchehr2020,gupta2022stochastic,pal2022inspection,pal2024random} for recent developments. For GBM specifically, resetting has already produced important results on stationary yet non-ergodic behavior, time averaging, and empirical applications \citep{stojkoski2021geometric,vinod2022nonergodicity,Vinod2022TimeAveraging,stojkoski2022income,Zanin2024SRGBM}. At the same time, resetting corresponds to a highly structured case in which exits and entries are tied together. The broader setting in which entries and exits occur separately, and potentially asymmetrically, remains much less explored.

This gap leaves several basic questions unresolved. Under what conditions does open multiplicative growth produce a stationary cross-sectional distribution? When does turnover suppress the explosive growth of moments, and when does it fail to do so? How do drift, volatility, entry, and exit jointly determine the shape of the stationary state? And how are first-passage properties altered once one moves beyond the balanced resetting case and allows recruitment and attrition to operate as distinct mechanisms? These questions are relevant not only for the theory of multiplicative stochastic processes, but also for the interpretation of empirical size distributions in systems where turnover is intrinsic rather than incidental.


To shed some light on these aspects, in this work, we study a geometric Brownian motion with constant entry and exit rates. At the level of the probability density, this leads to a Fokker-Planck equation with a source term at the entry point and a sink term representing exit. We solve this problem exactly and obtain the stationary distribution reached at long times. We show that this stationary state is a non-equilibrium steady state with asymmetric power-law tails around the entry point, with exponents determined by the drift, volatility, and the exit rate. We further derive closed-form expressions for the moments and show that they fall into three distinct regimes. They either saturate, grow linearly, or grow exponentially depending on whether exits dominates, balances, or is dominated by multiplicative growth. We also analyze first-passage properties in the presence of entries and exits, which allows us to characterize how turnover reshapes the time needed to reach a target value optimally and to compare these dynamics with the special balanced case of stochastic resetting.

Our results provide a minimal analytical framework for open multiplicative systems. In particular, they make explicit how entry, exit, drift, and volatility jointly determine stationary distributions, moment growth, and first-passage times. In this sense, the paper contributes to both the GBM literature and the broader literature on heavy-tailed size distributions by showing, in a transparent stochastic-process setting, how turnover transforms a standard proportional-growth model into an open non-equilibrium system with qualitatively new stationary and dynamical properties.


The paper is organized as follows. In Section~\ref{sec_formulation}, we introduce the GBM model under entry and exit mechanisms, formulate the renewal framework and present analytical results for the stationary state reached in the long time limit. We also present the temporal and stationary properties for the moments and log-moments. In Section~\ref{sec_fpt}, we investigate the first-passage properties of the system by calculating the survival probability and the MFPT. In Section \ref{sec_summary}, we discuss our findings, and conclude with future directions. Supporting calculations are presented in the Appendix.

\section{Geometric Brownian motion with entries and exits}\label{sec_formulation}

The geometric Brownian motion (GBM) is defined by the following Langevin equation \cite{black1973pricing}
\begin{align}\label{langevin eq}
dx(t)=\mu\,x(t)\,dt+\sigma\,x(t)\,dB(t), \quad x_0=x(0),
\end{align}
where $x(t)$ is the particle position. In practice, it could represent stock or asset price, fluctuating income, number of firms in a sector, etc. $x_0>0$ is the initial value of the observable, $\mu$ is the drift rate, $\sigma>0$ is the volatility, and $B(t)$ is the standard Brownian motion. Its solution in the It\^{o} interpretation of the multiplicative noise is given by
\begin{align}\label{langevin_eq_2}
    x(t)=x_0\,e^{(\mu-\sigma^2/2)t+\sigma\,B(t)}.
\end{align}
In other words, the GBM is a process in which the logarithm of $x(t)$ follows a Brownian motion with drift. This equation means that the incremental change $dx(t)$ in the asset price $x(t)$ at time $t$ is given by the deterministic aspects of the asset price, represented by the drift term $\mu\,x(t)\,dt$ and the stochastic aspects of the asset price dynamics, represented by the diffusion term $\sigma\,x(t)\,dB(t)$.  

The GBM can also be considered as a heterogeneous diffusion process~\cite{lau2007state,leibovich2019infinite} with a position-dependent diffusion coefficient $D(x)=\sigma^2x^2/2$ and external drift $\mu x(t)$. This follows from the fact that it can be rewritten in the following form
\begin{align}\label{langevin_eq}
\dot{x}(t)=\mu\,x(t)+\sigma\,x(t)\,\xi(t), \quad x_0=x(0),
\end{align}
where $\xi(t)=dB(t)/dt$ is white noise, which in this case is multiplicative. 

The corresponding Fokker-Planck equation in the It\^{o} interpretation --- which, for example, was used in the famed Black-Scholes-Merton model~\cite{black1973pricing,Merton1974} --- reads
\begin{align}\label{gbm_fpe}
    \frac{\partial}{\partial t}f_0(x,t)=-\mu\frac{\partial}{\partial x}xf_0(x,t)+\frac{\sigma^2}{2}\frac{\partial^2}{\partial x^2}x^{2}f_0(x,t),
\end{align}
with initial condition $f_0(x,t=0)=\delta(x-x_0)$. The solution of this equation is the log-normal distribution
\begin{align}\label{solution_GFP}
    f_0(x,t)=\frac{1}{x\sqrt{2\pi\sigma^2t}}\times\exp\left(-\frac{\left[(\log{x}-\log{x_0})-\bar{\mu} t\right]^{2}}{2\sigma^{2}t}\right),
\end{align}
where $\bar{\mu}=\mu-\sigma^2/2$. This solution means that the logarithm of the particle position, $\log x(t)$, is normally distributed with mean $\log x_0 + \bar{\mu} t$ and variance $\sigma^2 t$. As a consequence, $x(t)$ remains strictly positive for all times. This is a key feature that makes GBM well-suited for modeling quantities such as prices, firm sizes, or wealth. The effective drift $\bar{\mu} = \mu - \sigma^2/2$ reflects the well-known It\^{o} correction: even when the instantaneous drift $\mu$ is positive, multiplicative noise suppresses the typical trajectory relative to the mean, a manifestation of the non-ergodic character of GBM~\cite{peters2013ergodicity}. In particular, the mean $\langle x(t) \rangle = x_0 e^{\mu t}$ grows at rate $\mu$, while the median grows at the slower rate $\bar{\mu}$, so that the average is increasingly dominated by rare large trajectories as time progresses~\cite{peters2013ergodicity}.

Here, we consider the GBM model~(\ref{gbm_fpe}) with entries and exits, which is governed by the following Fokker-Planck equation
\begin{align}\label{gbm ito_fpe}
\frac{\partial}{\partial t}f(x,t)=&-\mu\frac{\partial}{\partial x}xf(x,t)+\frac{\sigma^2}{2}\frac{\partial^2}{\partial x^2}x^{2}f(x,t)\nonumber\\&-\lambda_m f(x,t)+\lambda_r\delta(x-x_0),
\end{align}
where $f(x,t)$ is the particle density, and $f(x,t=0)=\delta(x-x_0)$ is the initial condition. The third and fourth term from the right hand side of the equation are the loss of density due to exit at $x$ and the gain of density due to entries at $x_0$, and $\lambda_r$ and $\lambda_m$ are the entry and exit rates, respectively.

Equation~(\ref{gbm ito_fpe}) connects naturally to several related models in the literature. In the special case $\lambda_r=\lambda_m=r$, it reduces to the Fokker-Planck equation for GBM in the presence of Poissonian resetting~\cite{stojkoski2021geometric,stojkoski2022income,vinod2022nonergodicity,jolakoski2023first}, while the analogous problem (although restricted to first-passage) for standard diffusion with mortality and recruitment was recently studied in Refs.~\cite{linn2026dynamic,mercado2026stochastic}.

The parameters $\lambda_m$ and $\lambda_r$ admit natural interpretations in several empirical contexts: in firm dynamics they correspond to firm death and birth rates, respectively~\cite{Luttmer2007}, while in income dynamics they capture the case where the retirement and the first employment rates of individuals are not equal, contrary to the standard assumption where every ``retiring" individual is replaced with a new worker~\cite{gabaix2016dynamics}. As we will show, in both cases a stationary state emerges as a result of the interplay between entry and exit.

\subsection{Renewal formalism, probability density function and stationary state}

The solution of Eq.~\eqref{gbm ito_fpe} can be found by using the Laplace-Mellin transform, which is defined as 
\begin{align}
    \hat{\bar{f}}(q,s) = \int_0^\infty \left[\int_0^\infty e^{-st} f(x,t)\,dt\right]x^{q-1}\,dx.
\end{align}
By performing the Laplace transform with respect to $t$ and the Mellin transform with respect to $x$ in Eq. \eqref{gbm ito_fpe}, we obtain
\begin{align}\label{generalizedFPE_standard_app mellin}
\hat{\bar{f}}(q,s)=\frac{x_{0}^{q-1}(1+\lambda_r s^{-1})}{s+\lambda_m-\left[\frac{\sigma^2}{2}(q-1)(q-2)+\mu(q-1)\right]}.
\end{align}
The inverse Laplace transform of the above equation yields a renewal structure for the probability density function as follows
\begin{align}
    \bar{f}(q,t)=e^{-\lambda_m t}\bar{f}_0(q,t)+\lambda_r\int_{0}^{t}e^{-\lambda_m t'}\bar{f}_0(q,t')\,dt',
\end{align}
where
\begin{align}
    &\bar{f}_0(q,t)\nonumber\\&=x_{0}^{q-1}\times\exp\left(\frac{\sigma^2}{2}\left[q+\frac{1}{2}\left(\frac{2\mu}{\sigma^{2}}-3\right)\right]^2t-\frac{\bar{\mu}^2}{2\sigma^2}t\right),
\end{align}
is the solution of the Fokker-Planck equation~\eqref{gbm_fpe} for the GBM in Mellin space. Further doing an inverse Mellin transform we identify the renewal structure in the probability distribution function as follows  
\begin{align}\label{renew}
    f(x,t)=e^{-\lambda_m t}f_0(x,t)+\lambda_r\int_{0}^{t}e^{-\lambda_m t'}f_0(x,t')\,dt',
\end{align}
where $f_{0}(x,t)$ is the solution~(\ref{solution_GFP}) of the Fokker-Planck equation~(\ref{gbm_fpe}), \textit{i.e.}, the case with $\lambda_r=\lambda_m=0$. Note that for $\lambda_r=\lambda_m=r$, we recover the renewal equation for a stochastic process under Poissonian resetting~\cite{EvansMajumdar2011,pal2015diffusion,EvansMajumdarSchehr2020,maso2019transport,stojkoski2021geometric}
\begin{align}
    f_\text{reset}(x,t)=e^{-rt}f_0(x,t)+r\int_{0}^{t}e^{-rt'}f_0(x,t')\,dt'.
\end{align}
In the presence of entries and exit, the number of observed units is not conserved in each ensemble, which can be described using Eq.~\eqref{renew}. The normalization factor $\Phi(t)$, which is essentially the mean number of units at a given time $t$, can be determined by integrating Eq.~\eqref{renew} over space giving
\begin{align}\label{norm}
    \Phi(t) =& \int_0^\infty f(x,t) \,dx, \nonumber\\=&\dfrac{\lambda_r}{\lambda_m}+\left(1-\dfrac{\lambda_r}{\lambda_m}\right)e^{-\lambda_m t},
\end{align}
where we have used the normalization condition 
for the underlying probability density function (PDF) \textit{i.e.}, $\int_0^\infty f_0(x,t) dx=1$.

The mean number of particles in the system at a given time is independent of the spatial dynamics. To see this, we provide an alternative and general (dynamics independent) derivation of Eq.~\eqref{norm} in Appendix A. 

In the long time limit, $\Phi(t\to \infty)=\lambda_r/\lambda_m$ reaches a steady value that depends strongly on the rates of entries and exits.
This result also extends the stochastic-resetting framework~\cite{stojkoski2021geometric,vinod2022nonergodicity}, where the balanced condition $\lambda_r=\lambda_m$ constrains the steady-state population to unity, by allowing the long-run units count to take any positive value set freely by the entry-to-exit ratio. This generalization is empirically relevant in every open system discussed in the introduction, since entry and exit rates are almost never equal in firm markets, income cohorts, credit portfolios, etc.~\cite{DunneRobertsSamuelson1988,Hopenhayn1992,gabaix2016dynamics,dennis1991estimation,QuerciaStegman1992}

\begin{figure}[h!]
    \centering
    \includegraphics[width=\linewidth]{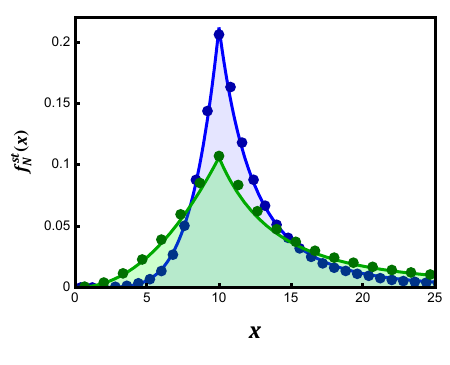}
    \caption{Stationary state $f^{st}_N(x)$ of GBM under entry and exit mechanism. Solid lines represent theoretical results for the steady state from Eq.~\eqref{st-det} with varying $\mu$ and $\sigma$ while keeping the initial condition $x_0=10$, entry rate $\lambda_r=100$ and exit rate $\lambda_m=0.1$ fixed. We set $(\mu,\sigma)=(0.02, \sqrt{0.01})$ for the plot colored in blue, and $(\mu,\sigma)$ to $(0.05,\sqrt{0.05})$ for the green curve. Markers indicate results from the numerical simulation demonstrating an excellent match.}
    \label{fig:ss}
\end{figure}

Taking $\Phi(t)$ into account, the normalized PDF can then be written as
\begin{align}\label{norm-pdf}
    f_N(x,t)=\dfrac{1}{\Phi(t)}\Big[&e^{-\lambda_m t}f_0(x,t)\nonumber\\&+\lambda_r\int_{0}^{t}e^{-\lambda_m u}f_0(x,u)\,du\Big].
\end{align} 
In the long time limit, from Eq.~(\ref{renew}), we obtain that the system approaches a stationary state with power-law tails
\begin{widetext}
\begin{align}
    \label{st-det}    f^{\text{st}}_N(x)&=\lim_{t\rightarrow\infty}f_N(x,t)=\dfrac{\lambda_r}{\lambda_r/\lambda_m}\int_{0}^{\infty}e^{-\lambda_m t'}f_0(x,t')\,dt'=\lambda_m\,\widetilde{f}_0(x,\lambda_m),\nonumber\\
    &=\frac{\lambda_m}{x\sqrt{\bar{\mu}^2+2\sigma^2\lambda_m}}\left\lbrace\begin{array}{lll}
    \left(\dfrac{x}{x_0}\right)^{-\frac{1}{\sigma^2}\left[\sqrt{\bar{\mu}^2+2\sigma^2\lambda_m}-\bar{\mu}\right]},     & x>x_0, \\ \\ 
    1 & x=x_0, \\ \\
    \left(\dfrac{x_0}{x}\right)^{-\frac{1}{\sigma^2}\left[\sqrt{\bar{\mu}^2+2\sigma^2\lambda_m}+\bar{\mu}\right]},     & x<x_0,
    \end{array}\right.
\end{align}
\end{widetext}
which has different behavior for $x>x_0$ and $x<x_0$. Here $\widetilde{f}_0(x,\lambda_m)=\mathcal{L}[f_0(x,t)](\lambda_m)$ denotes the Laplace transformation of $f_0(x,t)$ \textit{i.e.}, $\widetilde{f}_0(x,\lambda_m)=\int_0^\infty dt\, e^{-\lambda_m t}f_0(x,t)$. The stationary state for $\lambda_r=\lambda_m=r$ corresponds to the non-equilibrium stationary state obtained for the GBM with stochastic resetting~\cite{stojkoski2021geometric,stojkoski2022income,vinod2022time}, \textit{i.e.},
\begin{align}
    f_\text{reset}^{st}(x)=\frac{r}{x_0\sqrt{\bar{\mu}^2+2\sigma^2 r}}\left(\frac{x}{x_0}\right)^{\mp\frac{\sqrt{\bar{\mu}^2+2\sigma^2r}}{\sigma^2}+\frac{\mu}{\sigma^2}-\frac{3}{2}}.
\end{align}
The different behaviors of the stationary state for $x>x_0$ and $x<x_0$ result in a sharp cusp at the initial position $x=x_0$, which can be seen in Fig.~\ref{fig:ss}. 

Thus, it can be said that equation~\eqref{st-det} unifies, within a single exact GBM framework, distributional results that the existing literature has derived separately for individual applications. for example, the double-Pareto shape produced by birth--death Brownian motion~\cite{Reed2001,Reed2003,ReedJorgensen2004} emerges here as an exact consequence of the asymmetric entry and exit dynamics. Similarly, the tail exponent -- governed jointly by drift $\mu$, volatility $\sigma$, and attrition rate $\lambda_m$ -- recovers the Zipf-like upper tail of firm sizes~\cite{Axtell2001,Luttmer2007}, the Pareto tail of income and wealth distributions~\cite{gabaix2016dynamics,berman2020wealth,stojkoski2022ergodicity}, the heavy-tailed loan-balance distributions in credit markets~\cite{FooteGerardiWillen2018}, and the size distributions arising from proportional growth with turnover in ecological populations~\cite{lande2003stochastic,dennis1991estimation}. A key contribution beyond the existing literature is that Eq.~\eqref{st-det} makes the joint dependence of both tail exponents on $(\mu,\sigma,\lambda_m)$ analytically explicit, so that the effect of changing the mortality rate on the shape of the stationary distribution can be read off directly from the formula and tested across all these domains simultaneously.


\subsection{Analysis of the moments}
\begin{figure*}
    \centering
    \includegraphics[width=\linewidth]{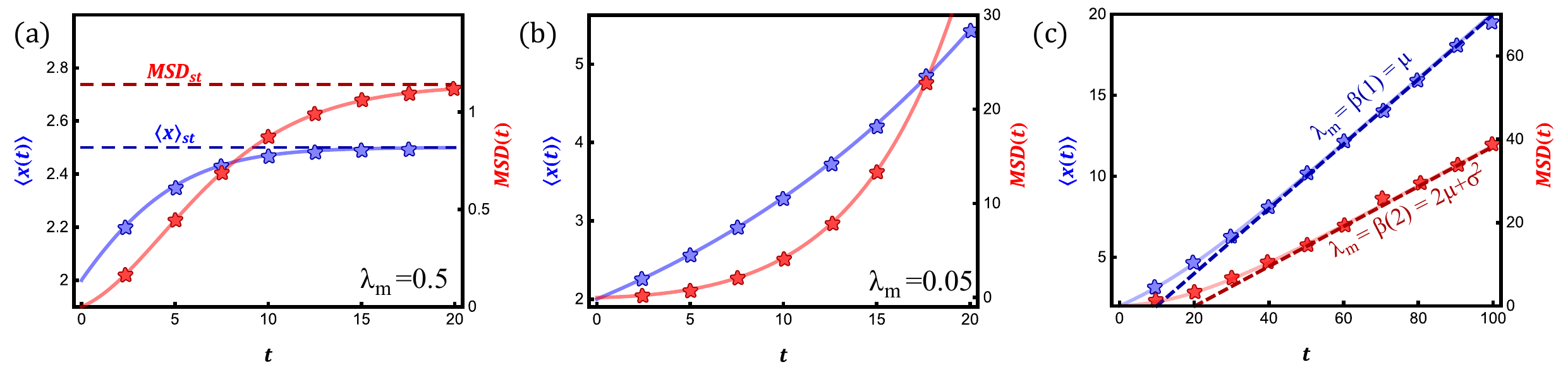}
    \caption{Temporal evolution of the mean and $\text{MSD}$ in the presence of entry and exit. Panel (a) indicates the initial exponential rise followed by a steady state convergence. Here, we have: rate of exit $\lambda_m=0.5$, $x_0=2$, $\mu=0.1$, $\sigma=\sqrt{0.02}$, and $\lambda_r=100$. Panel (b) describes the persistent exponential growth of the mean and $\text{MSD}$ with an exit rate $\lambda_m=0.05$. Panel (c) indicates the temporal behavior of the mean and $\text{MSD}$ over time, which is represented by solid dashed lines for $\lambda_m=\beta(1)=0.1$ and $\lambda_m=\beta(2)=0.12$. There, we see the linear growth in long time as predicted by the theoretical estimations (see Eqs.~\eqref{moments_lin} and \eqref{msd-large-t}) respectively. Across all the panels, solid lines denote the exact analytical results, while markers represent numerical estimations.}
    \label{fig:2}
\end{figure*}

From Eq.~\eqref{norm-pdf}, one can find the $n$-th moment of the variable $x$ by using the following relation 
\begin{align}\label{RM_x_moments}
   & \langle x^n(t) \rangle = \int_0^{\infty} x^n(t)\,f_N(x,t)\,dx \nonumber\\&= \dfrac{1}{\Phi(t)}\bigg[e^{-\lambda_m t}\langle x^n(t) \rangle_0
    +\lambda_r\int_{0}^{t}e^{-\lambda_m u}\langle x^n(t) \rangle_0\,du \bigg],
\end{align}
where $\langle x^n(t)\rangle_0$ represents the $n$-th moment of the position for the GBM ($\lambda_m=\lambda_r=0$). Using the following identity~\cite{stojkoski2021geometric}, $\langle (x(t))^n\rangle_0=x_0^n e^{\beta(n) t}$, with 
\begin{align}
 \beta(n) = n\mu +\frac{1}{2}n(n-1)\sigma^2,   
\label{beta}
\end{align}
for the underlying GBM process, one can write 
\begin{align}\label{mom-x}
    \langle x^n(t) \rangle = \dfrac{x_0^n}{\Phi(t)} \Big[&\dfrac{\lambda_r}{\lambda_m-\beta(n)}\nonumber\\&+\left(1-\dfrac{\lambda_r}{\lambda_m-\beta(n)}\right)e^{-(\lambda_m-\beta(n))t}\Big],
\end{align}
which is valid at all times. In particular, we observe three different regimes for the moments in the long time limit. For $\lambda_m>\beta(n)$ the moments saturate in the long time limit to the value 
\begin{align}\label{moments_sat}
   \langle x^n(t)\rangle\simeq x_0^n\frac{\lambda_m}{\lambda_m-\beta(n)}, 
\end{align}
see Table~\ref{tab-moments} for details. Note that with $n=1$, one can find the saturated value of mean, $\langle x\rangle_{st}=x_0\frac{\lambda_m}{\lambda_m-\mu}$ with $\lambda_m>\beta(1)$ as described in Fig.~\ref{fig:2}(a). For $\lambda_m<\beta(n)$ they have exponential dependence on time with $n$-dependent exponent, 
\begin{align}\label{moments_exp}
    \langle x^n(t)\rangle\simeq x_0^n\frac{\lambda_m}{\beta(n)-\lambda_m}e^{(\beta(n)-\lambda_m)t}.
\end{align}
In Fig.~\ref{fig:2}(b), we illustrate the exponential growth of the mean, which can be recovered from Eq.~\eqref{moments_exp} with $\lambda_m<\beta(1)$. While for $\lambda_m=\beta(n)$ the moments are linear in time with same slope, \textit{i.e.}, 
\begin{align}\label{moments_lin}
    \langle x^n(t)\rangle\simeq x_0^n\frac{\lambda_m}{\lambda_r}(1+\lambda_rt).
\end{align}
By setting $n=1$ in Eq.~\eqref{moments_lin}, we demonstrate the dependency of the mean on $t$ in Fig.~\ref{fig:2}(c).

Therefore, there is an interplay between the exit rate, volatility and drift in the dynamics of the moments. 

These three regimes extend both the standard closed GBM, which admits only exponential moment growth, and the resetting case $\lambda_r=\lambda_m$, which always yields moment saturation~\cite{stojkoski2021geometric}, by identifying the explicit condition $\lambda_m \gtrless \beta(n)$ as the boundary that governs moment behavior across open multiplicative systems. The phase boundary $\lambda_m = \beta(n) = n\mu + \tfrac{1}{2}n(n-1)\sigma^2$ could translate directly into empirical questions such as whether the exit rate in a given industry is sufficient to prevent runaway size concentration, or whether the mortality rate in a population is high enough to prevent unbounded growth of resource variance~\cite{Hopenhayn1992,stojkoski2022ergodicity,dennis1991estimation}. 

Moreover, the microscopic details can be effectively captured in terms of the mean-squared displacement (MSD), which reads
\begin{equation}\label{msd-x}
    \text{MSD}(t)=\langle (x(t)-x_0)^2\rangle,
\end{equation}
where the first two moments can be calculated from Eq.~\eqref{RM_x_moments} by putting $n=1$ and $2$, respectively. In Fig.~\ref{fig:2}, we provide the graphical demonstration of the behaviors of mean and MSD by considering two distinct regimes. Fig.~\ref{fig:2}(a) indicates the convergence to the following values $x_0\frac{\lambda_m}{\lambda_m-\mu}$ and $x_0^2\left[\frac{\lambda_m}{\lambda_m-(2\mu+\sigma^2)}-\frac{\lambda_m+\mu}{\lambda_m-\mu}\right]$ of mean and MSD, respectively.
On the other hand, the exponential growth can be observed 
in Fig.~\ref{fig:2}(b). In the short-time limit, both the mean and $\text{MSD}$ exhibit the linear temporal dependence; specifically, $\langle x(t)\rangle \simeq x_0+x_0\mu t$ and $\text{MSD}(t)\simeq x_0^2\sigma^2 t$. A similar growth can be observed with $\lambda_m\simeq \beta(2)$ in the large time limit for the mean (see Fig.~\ref{fig:2}(c)). Also, by utilizing Eq.~\eqref{mom-x} and Eq.~\eqref{msd-x}, we obtain
\begin{equation}\label{msd-large-t}
    \text{MSD}(t)\simeq x_0^2\left[\dfrac{3\lambda_m-\sigma^2}{\lambda_m+\sigma^2}+\dfrac{\lambda_m}{\lambda_r}(1+\lambda_r t)\right],
\end{equation}
the linear dependence of the MSD.  The variation of $\text{MSD}(t)$ in large $t$--limit is represented in Fig.~\ref{fig:2}(c).




\begin{table*}[ht]
    \centering
    \begin{tabular}{c|c|c}
        $n$-th moment & condition & value\\
        \hline
        $\langle x\rangle_\text{st}$ & $\lambda_m > \beta(1)(=\mu) $ & $x_0 \dfrac{\lambda_m}{\lambda_m-\mu}$\\
        \hline
        $\langle x^2\rangle_\text{st}$ & $\lambda_m > \beta(2)(=2\mu+\sigma^2) $ & $x^2_0 \dfrac{\lambda_m}{\lambda_m-(2\mu+\sigma^2)}$\\
        \hline
        \vdots & \vdots &\vdots \\
        \hline
        $\langle x^n\rangle_\text{st}$ & $\lambda_m > \beta(n)(=n\mu+\dfrac{1}{2}n(n-1)\sigma^2) $ & $x^n_0 \dfrac{\lambda_m}{\lambda_m-(n\mu+\dfrac{1}{2}n(n-1)\sigma^2)}$\\
    \end{tabular}
    \caption{Table with a consolidated list of the moments,  highlighting the conditions for convergence of the moments as $t\to\infty$, i.e., in the steady state.}
    \label{tab-moments}
\end{table*}



\subsection{Analysis of the log moments}
\begin{figure}
    \centering
    \includegraphics[width=\linewidth]{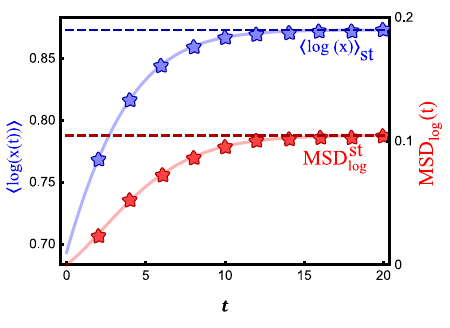}
    \caption{Transient behavior of log-mean (Eq.~\eqref{log-mean}) and mean squared displacement (MSD) in log-space (Eq.~\eqref{socond-log-moment}) as a function of time followed by the steady-state convergence for the parameters $x_0=2$, $\mu=0.1$, $\sigma=\sqrt{0.02}$, $\lambda_r=100$, and $\lambda_m=0.5$. Both curves approach a stationary value, as indicated by the dashed lines. Solid lines represent the theoretical predictions, while markers show the results from numerical simulations.}
    \label{fig:3}
\end{figure}

For economics applications, it is often useful to calculate the expectation of the periodic log returns, which is defined by
\begin{align}
    \frac{1}{\Delta t}\langle\log\left(x(t+\Delta t)/x(t)\right)\rangle\sim_{\Delta t\rightarrow0} \frac{d}{dt}\langle\log{x(t)}\rangle.
\end{align}
This is essentially the rate of first log-moment. The first log-moment (or log-mean) can be calculated by using the renewal structure as follows
\begin{align}
    \langle\log{x(t)}\rangle=\frac{1}{\Phi(t)}\Big[&e^{-\lambda_m t}\langle\log{x(t)}\rangle_0\nonumber\\&+\lambda_r\int_{0}^{t}e^{-\lambda_m u}\langle\log{x(t)}\rangle_0\,du\Big],
\end{align}
where $\langle\log{x(t)}\rangle_0=\log{x_0}+\bar{\mu}t$ is the log-mean of the GBM. Substituting the same in the renewal equation, we obtain 
\begin{align}\label{log-mean}
   & \langle\log{x(t)}\rangle=\frac{1}{\Phi(t)}\Big[e^{-\lambda_m t}\left(\log{x_0}+\bar{\mu}t\right)\nonumber\\&+\frac{\lambda_r}{\lambda_m}\log{x_0}\left(1-e^{-\lambda_m t}\right)+\frac{\lambda_r\bar{\mu}}{\lambda_m^2}\left(1-[\lambda_m t+1]e^{-\lambda_m t}\right)\Big],
\end{align}
which in the long-time limit ($t\rightarrow\infty$) saturates to the steady-state value
\begin{align}\label{log-mean-large}
\langle\log{x(t)}\rangle\sim\lambda_m\mathcal{L}\left[\langle\log{x(t)}\rangle_0\right](\lambda_m)=\log{x_0}+\bar{\mu}/\lambda_m.
\end{align} 
In Fig.~\ref{fig:3}, we demonstrate the behavior of the first log-moment (ref Eq.~\eqref{log-mean}), indicating its long time behavior, as mentioned in Eq.~\eqref{log-mean-large}. In a similar way, we can calculate the second log-moment,
\begin{align}
    \langle\log^2{x(t)}\rangle=\frac{1}{\Phi(t)}\Big[&e^{-\lambda_m t}\langle\log^2{x(t)}\rangle_0\nonumber\\&+\lambda_r\int_{0}^{t}e^{-\lambda_m u}\langle\log^2{x(t)}\rangle_0\,du\Big],
\end{align}
where $\langle\log^2{x(t)}\rangle_0=\log^2{x_0}+\left(2\bar{\mu}\log{x_0}+\sigma^2\right)t+\bar{\mu}^2t^2$ is the second log-moment of the GBM. Thus, the second log-moment reads
\begin{align}\label{socond-log-moment}
    \langle\log^2{x(t)}\rangle=&\Big[e^{-\lambda_m t}\left(\log^2{x_0}+\left(2\bar{\mu}\log{x_0}+\sigma^2\right)t+\bar{\mu}^2t^2\right)\nonumber\\&+\lambda_r\Big(\frac{\log^2{x_0}}{\lambda_m}[1-e^{-\lambda_m t}]\nonumber\\&+(2\bar{\mu}\log{x_0}+\sigma^2)\frac{1-e^{-\lambda_m t}(\lambda_m t+1)}{\lambda_m^2}\nonumber\\&+\bar{\mu}^2\frac{2-e^{-\lambda_m t}[2+\lambda_m t(\lambda_m t+2)]}{\lambda_m^3}\Big)\Big]/\Phi(t).
\end{align}
In the long-time limit, it approaches a constant value, as well,
\begin{align}
    \langle\log^2{x(t)}\rangle&\sim\mathcal{L}\left[\langle\log^2{x(t)}\rangle_0\right](\lambda_m)\nonumber\\&=\log^2{x_0}+\left(2\bar{\mu}\log{x_0}+\sigma^2\right)/\lambda_m+2\bar{\mu}^2/\lambda_m^2.
\end{align}
Thus, in the long-time limit, the variance becomes
\begin{align}
    \langle\log^2{x(t)}\rangle-\langle\log{x(t)}\rangle^2=\sigma^2/\lambda_m+\bar{\mu}^2/\lambda_m^2.
\end{align}
The MSD for the log-moments
MSD$_\text{log}(t) = \langle (\log(x(t))-\log(x_0))^2\rangle$ converges to $\sigma^2/\lambda_m+2\bar{\mu}^2/\lambda_m^2$ in the large time limit.
The graphical representation of the first moment (ref. Eq.~\eqref{log-mean}) and the MSD in log-space as a function of time is given in Fig.~\ref{fig:3}.

\subsection{Relaxation to the stationary state}

\begin{figure}
    \centering
    \includegraphics[width=\linewidth]{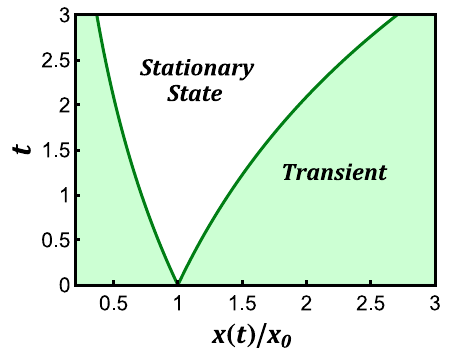}
    \caption{Growth of the boundary as a function of time separating the regions where the stationary state is achieved (central region) and the transient region where the system is yet to reach a steady state. We set $\sigma=1$ and $a=1$. Here, we have set $\mu=0.1,\sigma=\sqrt{0.02},~\&~\lambda_m=0.5$.}
    \label{fig:ness}
\end{figure}

We also analyze the behavior of the solution for large but finite time $t$. For large $t$, the second term in eq.~(\ref{renew}) is dominant, and thus we have
\begin{align}\label{renew_large_t}
    f(x,t)\approx \lambda_r\int_{0}^{t}e^{-\lambda_m t'}f_0(x,t')\,dt'.
\end{align}
We use the change in variables $\tau=wt$,
\begin{align}\label{renewal2}
f(x,t)\approx&\frac{\lambda_r\sqrt{t}}{x\sqrt{2\pi \sigma^2 }} \nonumber\\&\times\int_0^1~dw~\frac{e^{-rtw}}{\sqrt{w}}\exp \left( -\frac{\left[\log (x/x_0)-\bar{\mu}wt\right]^2}{2\sigma^2 wt}  \right).
\end{align}
From the Laplace approximation of the integral~\cite{arfken2005mathematical}, we find that the solution has the large deviation form
\begin{align}
f(x,t) \sim e^{-t I\left(\frac{\log(x/x_0)}{t}\right)},
\end{align}
where the large deviation function $I(y)$~\cite{majumdar2015dynamical} is given by
\begin{align}
\begin{array}{l}
I(y)=\left\{ \begin{array}{lll}
\sqrt{\frac{2a}{\sigma^2}}|y| &  & \text{for ~~}|y|<y^*\text{ ,}\\
a+\frac{y^2}{2 \sigma^2}&  & \text{for~~ }|y|>y^*\text{ ,}
\end{array}\right.\text{ }\end{array}
\label{LDF2}
\end{align}
where $y^*=\sqrt{2\sigma^2 a}$, $a=\lambda_m+\frac{\bar{\mu}^2}{2\sigma^2}$ and $y=\frac{\log\left(x/x_0)\right)}{t}$. Therefore, the boundary which separates the region in which the particle reaches the steady state from the transient region moves with non-constant velocity, since the stationary state is established in the inner core region given by
\begin{align}
x_0e^{-y^*t}<x(t)<x_0e^{y^*t}.
\end{align}
The boundary between these two regions is given in Fig.~\ref{fig:ness}.

\section{First-passage properties and optimal behavior}\label{sec_fpt}
A natural question in open multiplicative systems concerns how the observable $x(t)$; be it a firm size, an individual income, a loan balance, or a population count; reaches a prescribed threshold for the first time, and how the entry and exit mechanisms shape this threshold-crossing statistics. We frame this as a first-passage problem where the evolving quantity $x(t)$ is treated as a stochastic searcher, while the target value $x_T$ plays the role of an absorbing barrier. Exits from the system are modeled as a mortality mechanism for the searcher, while entries continuously replenish the pool of active searchers. A first-passage framework of this type was developed for GBM under stochastic resetting in~\cite{jolakoski2023first}; here we extend it to the general asymmetric case $\lambda_r \neq \lambda_m$. This mapping allows us to characterize how the interplay between multiplicative growth and turnover reshapes the time needed to reach a target, and to identify conditions under which the entry–exit asymmetry provides a speed-up relative to the symmetric resetting case.

Analysis of the first-passage properties for diffusive processes under entries and exits was recently done in~\cite{linn2026dynamic,mercado2026stochastic} following some earlier works~\cite{campos2024dynamic,tung2025first}. In this analysis, for the first-passage of GBM with entries and exits, we present a derivation and draw parallels to the approach used in~\cite{linn2026dynamic} along with \cite{mercado2026stochastic}. We assume that the process begins at $t=0$ with the deployment of a single searcher in the presence of a target. In this case, the survival probability that the target has not been found by the searcher up to time $t$ is given by
\begin{align}
    q_0(t)=\text{Prob}(T_0>t)&=\int_t^\infty d\tau ~\mathrm{P}_{T_0}(\tau),\nonumber\\
    &=1-\int_0^t d\tau\, \mathrm{P}_{T_0}(\tau),
\end{align}
where it has been assumed that the searcher does not exit and the corresponding first-passage time density is denoted by $\mathrm{P}_{T_0}(\tau)$. 

\begin{figure*}
    \centering
    \includegraphics[width=0.8\linewidth]{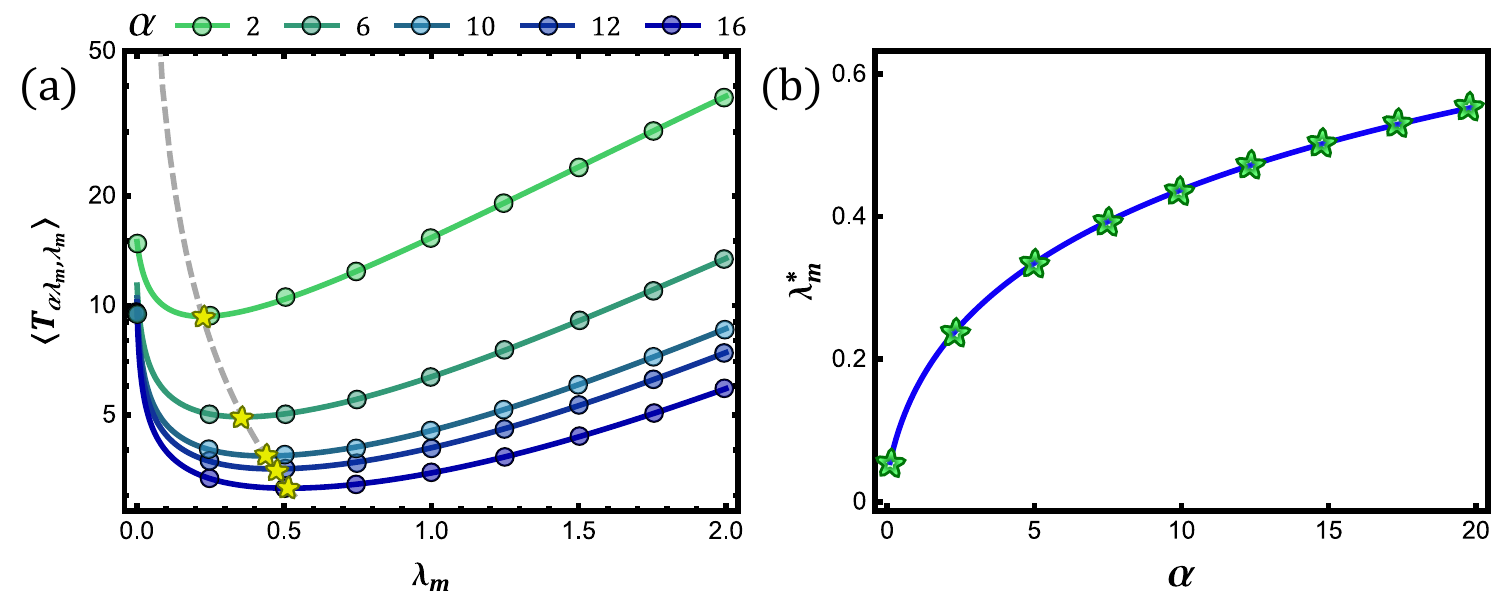}
    \caption{Panel (a): Variation of the MFPT (solid lines) with respect to the exit rate $\lambda_m$ for different $\alpha$. The minima of MFPTs are indicated with markers (yellow star), and the dashed line is the locus of the optimal exit rates $\lambda_m^*$. The solid lines are theoretical expressions (see Eq. (\ref{mfpt-death-birth})), verified against the markers from numerical simulations. Panel (b) demonstrates the variation of the optimal mortality rate (see Eq.~\eqref{trans-opt-lam}) with $\alpha$. Here, we have set the initial location $x_0=2$ with the following system parameters: $\mu=0.05, \sigma=\sqrt{0.02}$, while the threshold target is located at $x_T=3$.}
    \label{fig:5}
\end{figure*}

We now introduce a ``mortal'' searcher -- this means that the searcher can remain actively engaged until the target is found or it can also exit, i.e., abandon the search. A successful search can take place at time $\tau$ conditioned that the searcher remains active which occurs with the probability $e^{-\lambda_m \tau}$. Incorporating this fact into the above equation leads to the survival probability for the mortal searcher
\begin{equation}\label{mortal-surv}
    q_{\lambda_m}(t) = \text{Prob}(T_{\lambda_m}>t) =1-\int_0^t d\tau\, e^{-\lambda_m \tau}\mathrm{P}_{T_0}(\tau),
\end{equation}
from which one can identify the first-passage time density for the mortal searcher as 
\begin{equation}
    \mathrm{P}_{T_{\lambda_m}}(t) = -\dfrac{dq_{\lambda_m}}{dt} = e^{-\lambda_m t}\mathrm{P}_{T_0}(t).
\end{equation}
To model intermittent entry, we now define $q^{(n)}_{\lambda_m,\lambda_r}(t)$ as the joint survival probability given that exactly $n \geq 1$ additional searchers are introduced by time $t$. If a new searcher is deployed at some time $\tau > 0$, it only searches for a duration of $t-\tau$. Therefore, its individual survival probability is $q_{\lambda_m}(t-\tau)$. Assuming entry times $\tau$ are uniformly distributed over the interval $(0,t)$, the expected survival probability for any single new searcher is the time-average: $\frac{1}{t}\int_0^t d\tau\,q_{\lambda_m}(t-\tau)$, which simplifies to $\frac{1}{t}\int_0^t dt'\,q_{\lambda_m}(t')$. Since the searchers act independently and do not interact, the survival probabilities simply multiply. Hence, the probability that the initial searcher and all $n-1$ recruits fail to find the target by time $t$ is \cite{campos2024dynamic,tung2025first,linn2026dynamic}
\begin{equation}\label{rm-n}
    q^{(n)}_{\lambda_r,\lambda_m}(t) = q_{\lambda_m}(t)\left(\frac{1}{t}\int_0^t dt'\,q_{\lambda_m}(t')\right)^{n-1}.
\end{equation}
Next, we account for the random nature of the entry process. The number of new searchers arriving by time $t$ follows a Poisson distribution with a mean of $\lambda_r t$. By weighting the survival probability for $n$ searchers by the Poisson probability mass function, we obtain the total, unconditional survival probability $Q_{\lambda_m,\lambda_r}(t)$:
\begin{equation}\label{surv-rm-new}
    Q_{\lambda_r,\lambda_m}(t)=\sum_{n=1}^\infty \dfrac{e^{-\lambda_r t}\left(\lambda_r t\right)^{n-1}}{(n-1)!}q^{(n)}_{\lambda_r,\lambda_m}(t).
\end{equation}
Finally, substituting Eq.~\eqref{rm-n} into Eq.~\eqref{surv-rm-new} and evaluating the sum yields the closed-form expression for the survival probability~\cite{campos2024dynamic,grebenkov2020single,campos2024dynamic,tung2025first,tung2025passage,mercado2026stochastic,linn2026dynamic}
\begin{equation}
    Q_{\lambda_r,\lambda_m}(t)= q_{\lambda_m}(t) \exp\left(-\lambda_r \int_0^t d\tau \, \left[1-q_{\lambda_m} (\tau)\right]\right),
    \label{surv-rm-new-1}
\end{equation}
from which we can find the MFPT as follows
\begin{align}
\label{mfpt_form}
    \langle T_{\lambda_r,\lambda_m}\rangle &= \int_0^\infty dt~ \mathcal{Q}_{\lambda_r,\lambda_m}(t)\nonumber\\
    &=\int_0^\infty dt\,e^{-\lambda_r t} q_{\lambda_m}(t) \exp\left(\lambda_r \int_0^t d\tau \, q_{\lambda_m} (\tau)\right).
\end{align}
The above equation can be simplified further by denoting  $\mathcal{F}(t):=\exp\left(\lambda_r \int_0^t d\tau \, q_{\lambda_m} (\tau)\right)$ and further, applying integration by parts to the rest leading to
\begin{align}
    \langle T_{\lambda_r,\lambda_m}\rangle &= \int_0^tdt\,q_{\lambda_m}(t)e^{-\lambda_r t}\mathcal{F}(t)\nonumber\\
    &= \dfrac{1}{\lambda_r} \int_0^\infty dt\, e^{-\lambda_r t} \dfrac{d\mathcal{F}(t)}{dt}\nonumber\\
    &= \int_0^\infty dt\,e^{-\lambda_r t}\mathcal{F}(t)-\dfrac{1}{\lambda_r}.\nonumber
\end{align}
Finally substituting back $\mathcal{F}(t)=\exp\left(\lambda_r \int_0^t d\tau \, q_{\lambda_m} (\tau)\right)$ and further simplifications, we arrive at
\begin{align}\label{mfpt-death-birth}
    \langle T_{\lambda_r,\lambda_m}\rangle=&\int_0^\infty dt\, \exp\bigg(-\lambda_r \int_0^t d\tau\, e^{-\lambda_m \tau}(t-\tau)\mathrm{P}_0(\tau)\bigg)\nonumber\\&-\dfrac{1}{\lambda_r}.
\end{align}
To delve deeper into the role of entry and exit mechanisms, it is useful to define a relative ratio 
\begin{equation}
 \lambda_r=\alpha\lambda_m,   
\end{equation}
and substitute in Eq.~\ref{mfpt-death-birth}). 
To find the optimal exit rate $\lambda_m^*$, we use the minimization condition
\begin{equation}
    \left.\dfrac{d\langle T_{\alpha\lambda_m,\lambda_m}\rangle}{d \lambda_m}\right|_{\lambda_m=\lambda_m^*}=0,
\end{equation}
where
\begin{align}
    \dfrac{d\langle T_{\alpha\lambda_m,\lambda_m}\rangle}{d \lambda_m}=&-\alpha\int_0^\infty dt\, e^{-\alpha\lambda_m g(\lambda_m,t)}\nonumber\\&\times\left[g(\lambda_m,t)+\lambda_m \dfrac{dg(\lambda_m,t)}{d\lambda_m}\right]+\dfrac{1}{\alpha \lambda_m^2},
\end{align}
and we further have defined $g(\lambda_m,t)=\int_0^t d\tau\, e^{-\lambda_m \tau}(t-\tau) \mathrm{P}_0(\tau)$. The optimal exit rate can then be obtained from the following transcendental equation 
\begin{align}\label{trans-opt-lam}
    \dfrac{1}{(\lambda_m^*)^2}=\alpha^2 \int_0^\infty &dt\, e^{-\alpha\lambda_m g(\lambda_m,t)}\nonumber\\&\left.\times\left[g(\lambda_m,t)+\lambda_m \dfrac{dg(\lambda_m,t)}{d\lambda_m}\right]\right|_{\lambda_m=\lambda_m^*}.
\end{align}
The existence of $\lambda_m^*$ generalizes the optimal resetting results of Refs.~\cite{EvansMajumdar2011,PalReuveni2017}, which are confined to the symmetric case $\alpha=1$, to the full range of asymmetric regimes $\alpha=\lambda_r/\lambda_m\neq 1$ that characterize many empirical open systems, including firm markets where entry typically exceeds exit, income and wealth processes where births and deaths operate at different rates, credit portfolios where origination and default rates differ, and ecological populations where recruitment and mortality are independently governed~\cite{DunneRobertsSamuelson1988,gabaix2016dynamics,QuerciaStegman1992,lande2003stochastic}. In Fig. ~\ref{fig:5}, we present the dependence of the MFPT with respect to the exit rate for different values of $\alpha=\lambda_r/\lambda_m$, and fixed $\alpha$, respectively. We also show the behavior of the optimal exit rates for various $\alpha$ which indicates a non-trivial optimization in this process.

\subsection*{A speed-up comparison between entry-exit and stochastic resetting}
As mentioned earlier, the entry and exit dynamics constitute a fundamentally non-conservative process, since the total number of particles evolves in time and the normalization factor $\Phi(t)$ depends explicitly on the entry and exit rates. This is in sharp contrast to stochastic resetting, where the number of particles is conserved and resetting merely redistributes probability within the system without altering the total mass. Consequently, although both mechanisms can generate stationary states and accelerate first-passage events, the physical origins of these effects are markedly different.

\begin{figure}
    \centering
    \includegraphics[width=\linewidth]{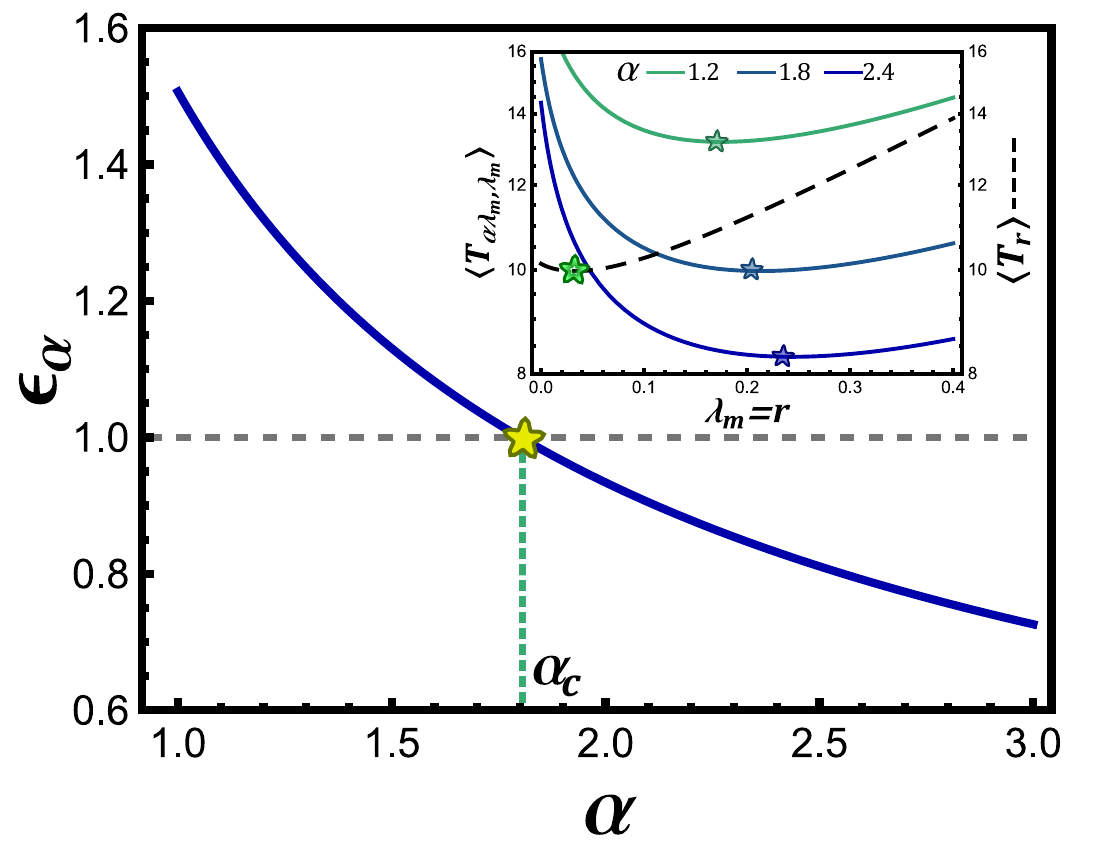}
    \caption{Scanning the behavior of the speed-up ratio $\epsilon_{\alpha}$ as a function of $\alpha$. In the domain with $\epsilon_\alpha>1$, where stochastic resetting outpaces the entry-exit process. Beyond $\alpha>\alpha_c$, we observe an alternative scenario (with $\epsilon_\alpha<1$) in which the optimally conducted entry-exit process becomes more efficient in optimizing the MFPT, as can also be seen in the inset plot where we have plotted the MFPTs for respective cases and marked the optimal points with star markers. We found that the critical value of the ratio between recruitment and mortality rate is given by $\alpha_c\approx 1.8$ for the following set of parameters $x_0=2,x_T=3,\mu=0.05~\&~\sigma=\sqrt{0.02}$.}
    \label{fig:4}
\end{figure}

In the entry and exit framework, the enhancement of first-passage properties arises from the combined effect of searcher injection and removal, which continuously reshapes the underlying population dynamics. By contrast, stochastic resetting expedites search by interrupting long unfavorable excursions and repeatedly redirecting trajectories toward the initial configuration. It is therefore important to quantitatively compare the relative speed-up achieved under these two mechanisms when applied independently to the same first-passage process. To this end, we consider the ratio 
\begin{equation}\label{trade_off_opt}
    \epsilon_\alpha = \dfrac{\langle T_{\alpha \lambda_m^*, \lambda_m^*}\rangle}{\langle T_{r^*}\rangle},
\end{equation}
where both the MFPTs in the numerator and denominator are respectively optimized. The numerator is given by Eq.~(\ref{mfpt-death-birth}) but optimized at $\lambda_m^*$. On the other hand, the MFPT in the presence of stochastic resetting at a rate $r$ has been studied extensively over the years and can be written as \cite{reuveni2016optimal,PalReuveni2017} 
\begin{equation}\label{mfpt_res_def}
    \langle T_r\rangle = \dfrac{1-\widetilde{T}_0(r)}{r\widetilde{T}_0(r)},
\end{equation}
where $\widetilde{T}_0(r)=\int_0^\infty dt ~e^{-rt}\mathrm{P}_{T_0}(t)$ is the Laplace transform of the first-passage time density $\mathrm{P}_{T_0}(t)$ of the GBM process in the absence of resetting and is given by \cite{jolakoski2023first}
\begin{equation}
    \widetilde{T}_0(r)=\left(\dfrac{x_T}{x_0}\right)^{\dfrac{\bar{\mu}-\sqrt{\bar{\mu}^2+2 r \sigma ^2}}{\sigma ^2}}.
\end{equation}
Incorporating the above in Eq.~\eqref{mfpt_res_def} yields the MFPT of GBM under stochastic resetting. 
The speed-up ratio $\epsilon_\alpha$ 
for a fixed initial 
$x_0$ and target $x_T$, is plotted 
in Fig.~\ref{fig:4} which shows the range of $\alpha$ for which stochastic resetting expedites the entry and exit model and vice-versa. In particular, the result that the entry and exit process outperforms optimal resetting for a range of $\alpha>\alpha_c$, where $\alpha_c$ is a critical value that depends on the system parameters \& entry-exit and resetting rates, shows that allowing entry and exit to operate asymmetrically is not merely a technical generalization but can qualitatively improve first-passage efficiency. This finding was inaccessible within the balanced resetting framework and carries direct implications for understanding optimal turnover across all these domains.


\section{Conclusion}\label{sec_summary}
Most analytical treatments of GBM describe closed systems in which the number of trajectories is fixed. Many empirically relevant systems are, however, open. Firms continuously enter and exit markets~\cite{DunneRobertsSamuelson1988,Hopenhayn1992}, incomes are generated and extinguished at unequal rates~\cite{gabaix2016dynamics,berman2020wealth}, loans are originated and defaulted independently~\cite{QuerciaStegman1992,FooteGerardiWillen2018}, and biological populations are governed by separate birth and death processes~\cite{dennis1991estimation,lande2003stochastic}. The framework of stochastic resetting~\cite{EvansMajumdar2011,PalReuveni2017} is the closest existing approach but imposes $\lambda_r=\lambda_m$, ruling out the asymmetry that is typical of all these settings. The present work removes this constraint and treats $\lambda_r$ and $\lambda_m$ as fully independent parameters.

The analysis yields four results. First, the system always relaxes to a non-equilibrium stationary state whose double-power-law shape is governed by $(\mu,\sigma,\lambda_m)$ through Eq.~\eqref{st-det}, with long-run population size $\Phi(\infty)=\lambda_r/\lambda_m$, which is a free parameter absent in the resetting case. Second, the $n$-th moment exhibits three distinct regimes: saturation, linear growth, or exponential divergence. These are separated by the phase boundary $\lambda_m=\beta(n)=n\mu+\frac{1}{2}n(n-1)\sigma^2$ (Eqs.~\eqref{moments_sat}--\eqref{moments_exp}). Third, relaxation to the stationary state proceeds through an inner core region whose boundary expands with velocity $y^*=\sqrt{2\sigma^2(\lambda_m+\bar{\mu}^2/2\sigma^2)}$. Finally, the MFPT is minimized at an optimal exit rate $\lambda_m^*$, and the RM process outperforms optimal stochastic resetting whenever $\alpha=\lambda_r/\lambda_m>1$.

These results have implications at both theoretical and empirical levels. On the one hand, stochastic process theory, the ratio $\alpha$ emerges as a single control parameter that governs population size, distribution shape, moment behavior, and first-passage efficiency simultaneously, with stochastic resetting recovered as the special case $\alpha=1$. On the other hand, for applications, Eq.~\eqref{st-det} provides an analytical expression for double-Pareto size distributions observed in firms~\cite{Axtell2001,Luttmer2007}, incomes and wealth~\cite{gabaix2016dynamics,stojkoski2022ergodicity}, and ecological populations~\cite{lande2003stochastic,dennis1991estimation}, with both tail exponents shifting analytically as $\lambda_m$ changes. The moment phase boundary translates directly into an empirically testable condition: turnover must exceed $\beta(n)$ for the $n$-th moment to remain bounded, a question relevant to concentration dynamics in industries~\cite{Hopenhayn1992} and inequality in income distributions~\cite{gabaix2016dynamics}.

But, our results are not without limitations. First, the model rests on several idealizations that might not be practical. For instance, both rates are constant and state-independent, whereas real exit rates often depend on firm age or size and entry rates vary cyclically~\cite{Hopenhayn1992}. Similarly, all recruits enter at the same point $x_0$, whereas real entrants are heterogeneous; a distribution of entry sizes would modify the stationary shape. On the analysis side, our findings are exact within the It\^{o} interpretation. The Stratonovich and H\"{a}nggi-Klimontovich cases can be derived and a systematic empirical comparison of the three interpretations is left for future works.

Despite these simplifications, the entry and exit framework provides a minimal, analytically tractable foundation for open multiplicative systems. The key insight is that the asymmetry between entry and exit reshapes the dynamics, and in particular, the stationary distribution, the moment evolution, and the efficiency of reaching threshold. It would be interesting to extend this framework to more realistic economic and financial settings, including heterogeneous agents, interacting markets, and time-dependent recruitment or attrition policies. Calibrating the predicted tail exponents and optimal turnover rates against longitudinal data on firm sizes, income dynamics, and ecological populations is also a natural and promising direction for future work.

\section{Data Availability}
All the data are available within the main text.

\begin{acknowledgments}
SP is grateful to the Indian Statistical Institute, Kolkata, for the Research Fellowship support. AP gratefully acknowledges research support from the Department of Atomic Energy, Government of India.  AP acknowledges research funding under the scheme ANRF/ARGM/2025/001623 from ANRF, India. TS is supported by the German Science Foundation (DFG, Grant number ME 1535/12-1) and by the Alliance of International Science Organizations (Project No. ANSO-CR-PP2022-05). TS was also supported by the Alexander von Humboldt Foundation.
\end{acknowledgments}

\newpage
\onecolumngrid
\appendix

\section{Alternative derivation of $\Phi(t)$ -- Eq. (\ref{norm})}
Let \(P_n(t)\) denote the probability of having \(n\) particles at time \(t\). We assume that an entry (recruitment/birth) occurs with rate \(\lambda_r\), while exit (loss/death) occurs with rate \(\lambda_m\) per particle. The corresponding entry-exit process is given by
\begin{equation}
    n \xrightarrow{\lambda_r} n+1,\qquad n \xrightarrow{n \lambda_m } n-1.
\end{equation}

The master equation governing the evolution of \(P_n(t)\) reads
\begin{equation}
    \frac{dP_n(t)}{dt}=\lambda_r P_{n-1}(t)+\lambda_m (n+1) P_{n+1}(t)-(\lambda_r+\lambda_m n)P_n(t),\qquad n\ge 0,
\end{equation}
with the convention
\begin{equation}
    P_{-1}(t)=0.
\end{equation}

The different terms in the master equation have the following interpretation:

\begin{itemize}
\item \(\lambda_r P_{n-1}(t)\): gain into state \(n\) due to entry from state \(n-1\),
\item \(\lambda_m (n+1)P_{n+1}(t)\): gain into state \(n\) due to exit from state \(n+1\),
\item \((\lambda_r+\lambda_m n)P_n(t)\): loss from state \(n\) due to either entry or exit.
\end{itemize}

Given this, we can now define the mean particle number in the system as
\begin{equation}
    \langle n(t)\rangle=\sum_{n=0}^{\infty} nP_n(t).
\end{equation}

To derive its evolution equation, we multiply the master equation by \(n\) and sum over all \(n\):
\begin{equation}
    \frac{d}{dt}\langle n\rangle=\sum_{n=0}^\infty n\frac{dP_n}{dt}.
\end{equation}

Substituting the master equation, we obtain
\begin{align}
    \frac{d\langle n\rangle}{dt}&=\lambda_r \sum_n nP_{n-1}+\lambda_m\sum_n n(n+1)P_{n+1}-\lambda_r\sum_n nP_n-\lambda_m\sum_n n^2P_n.
\end{align}

Now shifting indices appropriately,
\begin{equation}
    \sum_n nP_{n-1}=\sum_m (m+1)P_m=\langle n\rangle+1,
\end{equation}
and
\begin{equation}
    \sum_n n(n+1)P_{n+1} = \sum_m (m-1)mP_m =\langle n^2\rangle-\langle n\rangle.
\end{equation}

Substituting these expressions back yields
\begin{align}
    \frac{d\langle n\rangle}{dt}&=\lambda_r(\langle n\rangle+1)+\lambda_m(\langle n^2\rangle-\langle n\rangle)-\lambda_r\langle n\rangle-\lambda_m\langle n^2\rangle.
\end{align}

The second moments cancel exactly, leading to the closed equation
\begin{equation}
    \boxed{\frac{d\langle n\rangle}{dt}=\lambda_r-\lambda_m\langle n\rangle.}
\end{equation}

Solving this equation with the initial condition
\begin{equation}
    \langle n(0)\rangle=n_0,
\end{equation}
we obtain
\begin{equation}\label{app-gen}
    \boxed{\Phi(t)=\langle n(t)\rangle=\frac{\lambda_r}{\lambda_m}+\left(n_0-\frac{\lambda_r}{\lambda_m}\right)e^{-\lambda_m t}.}
\end{equation}
Since we initiate the process with one entry so that $n_0=1$, from Eq.~\eqref{app-gen}, we recover
\begin{equation}\label{app-phi}
    \boxed{\Phi(t)=\langle n(t)\rangle=\frac{\lambda_r}{\lambda_m}+\left(1-\frac{\lambda_r}{\lambda_m}\right)e^{-\lambda_m t},}
\end{equation}
which was announced in Eq.~\eqref{norm} in the main text. Thus, the mean population relaxes exponentially toward the stationary value
\begin{equation}
    \boxed{\Phi(t \to \infty)=\langle n\rangle_{\rm st}=\frac{\lambda_r}{\lambda_m},}
\end{equation}
which was also mentioned in the main text.

\section{First-passage time density of the underlying GBM process}
\begin{figure}[t]
    \centering
    \includegraphics[width=0.8\linewidth]{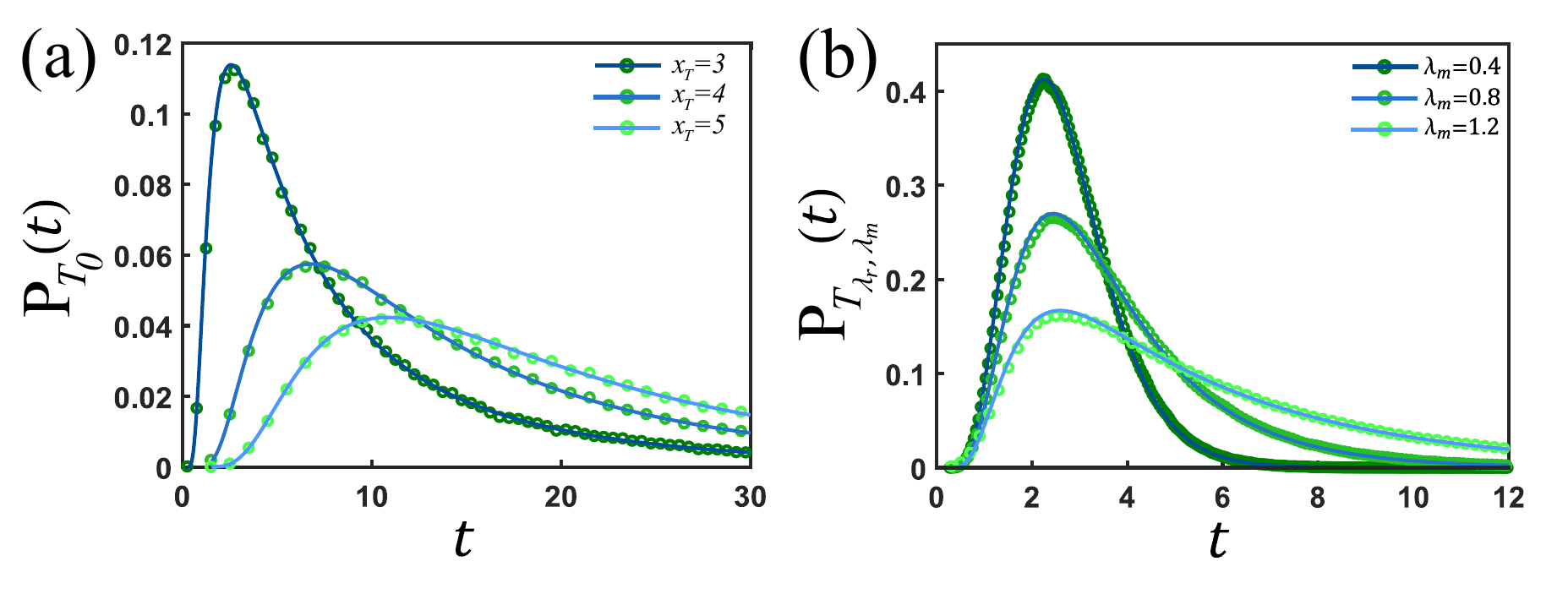}
    \caption{Panel (a) illustrates the FPT density of the underlying GBM process (see Eq.~\eqref{fpt-den}), with different targets $x_T=3,4,~\&~5$. Other system parameters are fixed at $\mu=0.05,\sigma=\sqrt{0.02}$ and the initial condition $x_0=2$. Theoretical analysis is represented with solid lines, and markers indicate the numerical estimation. In contrast to this, we describe the FPT density of GBM in the presence of entry and exit mechanism (see Eq.~\eqref{fpt-RM}) with rates $\lambda_r, \lambda_m$, respectively, in panel (b). By keeping fixed $\lambda_r=10$, we vary the exit rate $\lambda_m=0.4,0.8,1.2$, as indicated in panel(b).}
    \label{fig-fpt-RM}
\end{figure}

In this section, we provide the FPT density in the presence of an absorbing target $x_T>0$. To this end, we start by recalling the governing stochastic equation for the GBM from the main text (see Eq.~\eqref{langevin_eq})
\begin{equation}
    dx(t) = \mu x(t) dt+\sigma x(t) dB(t),
\end{equation}
and consider the following transformation $y(x(t))=\log(x(t))$. Now, using It\^{o}'s lemma, we get
\begin{equation}\label{ito_1}
    dy(x(t)) = y'(x(t)) dx(t)+\frac{1}{2}y''(x(t))(dx(t))^2,
\end{equation}
further substituting $y'(x(t))=\frac{1}{x(t)},y''(x(t))=\frac{1}{(x(t))^2}$ and $(dx(t))^2=(x(t))^2\sigma^2 dt$, from the above identity, we get
\begin{equation}\label{ito_2}
    dy(x(t)) = \frac{1}{x(t)}\left(\mu x(t) dt+\sigma x(t) dB(t)\right)-\frac{\sigma^2}{2} dt.
\end{equation}
Upon rearranging, we finally arrive at
\begin{equation}\label{ito_3}
    dy(x(t)) = \left(\mu-\frac{1}{2}\sigma^2\right) dt + \sigma dB(t),
\end{equation}
which we identify as a drift-diffusion process in $y$-space
\begin{equation}\label{langevin_log}
    dy = \bar{\mu} dt + \sigma dB(t),
\end{equation}
with the initial condition $y_0=\log x_0$. One can rewrite the PDF of this underlying process, in the absence of any target, in the following way
\begin{equation}\label{pdf-wo-log}
    g_0(y,t)=f_0(x,t)\left|\dfrac{dx}{dy}\right|=\dfrac{1}{\sqrt{2\pi\sigma^2 t}}\exp\left(-\dfrac{(y-y_0-\bar{\mu}t)^2}{2\sigma^2 t}\right).
\end{equation}
where recall that $f_0(x,t)$ denotes the PDF of the underlying process (Eq.~\eqref{solution_GFP}), whereas $g_0(y,t)$ represents the PDF in log-space. 
In the presence of a target at $y_T=\log x_T$, utilizing the image method, the PDF can be easily computed in the log space as
\begin{equation}\label{image-pdf-wo-log}
    \mathcal{G}_0(y,t)=\dfrac{1}{\sqrt{2\pi \sigma^2 t}}\left[\exp\left(-\dfrac{(y-y_0-\bar{\mu}t)^2}{2\sigma^2 t}\right)-\exp\left(\dfrac{2\bar{\mu}}{\sigma^2}(y_T-y_0)\right)\exp\left(-\dfrac{(y-(2y_T-y_0)-\bar{\mu}t)^2}{2\sigma^2 t}\right)\right].
\end{equation}
Furthermore, utilizing Eq.~\eqref{image-pdf-wo-log}, the survival probability reads
\begin{equation}\label{image-surv-wo-log}
    q_0(t) = \int dy ~\mathcal{G}_0(y,t)= \dfrac{1}{2}\left[\text{Erf} \left(\dfrac{\bar{\mu}t-(y_T-y_0)}{\sqrt{2\sigma^2 t}}\right)-\exp\left(\dfrac{2\bar{\mu}}{\sigma^2}(y_T-y_0)\right)\text{Erf} \left(\dfrac{\bar{\mu}t+(y_T-y_0)}{\sqrt{2\sigma^2 t}}\right)\right].
\end{equation}
The FPT density can be written as
\begin{equation}
    \mathrm{P}_{T_0}(t) =-\partial_t q_0(t)=\dfrac{|y_T-y_0|}{\sqrt{2\pi \sigma^2 t^3}}\exp\left(-\dfrac{(y_T-y_0-\bar{\mu}t)^2}{2 \sigma^2 t}\right).
\end{equation}
Finally, putting $y_T=\log{x_T}$ and $y_0=\log{x_0}$, we recover the FPT density in the $x$-space
\begin{equation}\label{fpt-den}
    \mathrm{P}_{T_0}(t) = \dfrac{|\log(x_T/x_0)|}{\sqrt{2\pi \sigma^2 t^3}}\exp\left(-\dfrac{(\log(x_T/x_0)-\bar{\mu}t)^2}{2 \sigma^2 t}\right).
\end{equation}
We plot Eq.~\eqref{fpt-den} in Fig.~\ref{fig-fpt-RM}(a) against the numerical simulations which matches perfectly. 
From Eq.~\eqref{mortal-surv}, the FPT density in the presence of mortality with rate $\lambda_m$ yields
\begin{equation}
    \mathrm{P}_{T_{\lambda_m}}(t) = -\dfrac{dq_{\lambda_m}}{dt} = e^{-\lambda_m t}\mathrm{P}_{T_0}(t).
\end{equation}

\section{First-passage time density under entry and exit mechanism}
We recall Eq. \eqref{surv-rm-new-1}) for the survival probability from the main text
\begin{equation}\label{q1-surv}
    Q_{\lambda_r,\lambda_m}(t)=e^{-\lambda_r t} q_{\lambda_m}(t) \exp\left(\lambda_r \int_0^t d\tau \, q_{\lambda_m} (\tau)\right).
\end{equation}
Using Eq.~\eqref{mortal-surv}, one can rewrite the  Eq.~\eqref{q1-surv} in the following way
\begin{equation}
    Q_{\lambda_r,\lambda_m}(t)=q_{\lambda_m} (t)\exp\left(-\lambda_r\int_0^td\tau\,\int_0^\tau du\, e^{-\lambda_m u}P_0(u)\right).
\end{equation}
Finally, using $\mathrm{P}_{T_{\lambda_r,\lambda_m}}(t)=-\partial_t Q_{\lambda_r,\lambda_m}(t)$, the FPT density under entry-exit mechanism can be found as
\begin{align}
    \mathrm{P}_{T_{\lambda_r,\lambda_m}}(t) &=-\partial_t Q_{\lambda_r,\lambda_m}(t) \nonumber\\
    &= \exp\left(-\lambda_r\int_0^td\tau\,\int_0^\tau du\, e^{-\lambda_m u}\mathrm{P}_0(u)\right) \left(e^{-\lambda_m t} \mathrm{P}_0(t)+\lambda_r \left(1-\int_0^t e^{-\lambda_m \tau}\mathrm{P}_0(\tau)\,d\tau\right)\int_0^t e^{-\lambda_m \tau}\mathrm{P}_0(\tau)\,d\tau\right),
\end{align}
which can be rewritten in the following way 
\begin{equation}\label{fpt-RM}
    \mathrm{P}_{T_{\lambda_r,\lambda_m}}(t) = \exp\left(-\lambda_r\int_0^td\tau\,\int_0^\tau du\, \mathrm{P}_{\lambda_m}(u)\right)\bigg[\mathrm{P}_{\lambda_m}(t)+\lambda_r \left(1-q_{\lambda_m}(t)\right)q_{\lambda_m}(t)\bigg].
\end{equation}
In Fig.~\ref{fig-fpt-RM}(b), we illustrate the behavior of the FPT density under entry-exit mechanism (as indicated in Eq.~\eqref{fpt-RM}) against numerical simulations for different exit rates.

\bibliography{apssamp}

\end{document}